\documentclass[aps,prl,twocolumn,superscriptaddress,floatfix]{revtex4-1}
\usepackage{graphicx}
\usepackage[caption=false]{subfig}
\usepackage{braket}
\usepackage{mathtools}
\usepackage{color}
\graphicspath{{figures/}}

\begin{document}

% Use the \preprint command to place your local institutional report
% number in the upper righthand corner of the title page in preprint mode.
% Multiple \preprint commands are allowed.
% Use the 'preprintnumbers' class option to override journal defaults
% to display numbers if necessary
%\preprint{}

%Title of paper
\title{Amplitude sensing below the zero-point fluctuations with a two-dimensional trapped-ion mechanical oscillator}

% repeat the \author .. \affiliation  etc. as needed
% \email, \thanks, \homepage, \altaffiliation all apply to the current
% author. Explanatory text should go in the []'s, actual e-mail
% address or url should go in the {}'s for \email and \homepage.
% Please use the appropriate macro foreach each type of information

% \affiliation command applies to all authors since the last
% \affiliation command. The \affiliation command should follow the
% other information
% \affiliation can be followed by \email, \homepage, \thanks as well.
\author{K. A. Gilmore}
\email[]{kevin.gilmore@colorado.edu}
%\homepage[]{Your web page}
%\thanks{}
%\altaffiliation{}
\affiliation{National Institute of Standards and Technology, Boulder, Colorado 80305, USA}
\affiliation{JILA and Department of Physics, University of Colorado, Boulder, Colorado, 80309, USA}

\author{J. G. Bohnet}
\affiliation{National Institute of Standards and Technology, Boulder, Colorado 80305, USA}

\author{B. C. Sawyer}
\affiliation{Georgia Tech Research Institute, Atlanta, Georgia 30332, USA}

\author{J. W. Britton}
\affiliation{U.S. Army Research Laboratory, Adelphi, Maryland 20783, USA}

\author{J. J. Bollinger}
\email[]{john.bollinger@nist.gov}
\affiliation{National Institute of Standards and Technology, Boulder, Colorado 80305, USA}

%Collaboration name if desired (requires use of superscriptaddress
%option in \documentclass). \noaffiliation is required (may also be
%used with the \author command).
%\collaboration can be followed by \email, \homepage, \thanks as well.
%\collaboration{}
%\noaffiliation

\date{\today}

\begin{abstract}
We present a technique to measure the amplitude of a center-of-mass (COM) motion of a two-dimensional ion crystal of $\sim$100 ions. By sensing motion at frequencies far from the COM resonance frequency, we experimentally determine the technique's measurement imprecision. We resolve amplitudes as small as 50 pm, 40 times smaller than the COM mode zero-point fluctuations. The technique employs a spin-dependent, optical-dipole force to couple the mechanical oscillation to the electron spins of the trapped ions, enabling a measurement of one quadrature of the COM motion through a readout of the spin state. We demonstrate sensitivity limits set by spin projection noise and spin decoherence due to off-resonant light scattering. When performed on resonance with the COM mode frequency, the technique demonstrated here can enable the detection of extremely weak forces ($< \,$1 yN) and electric fields ($< \,$1 nV/m), providing an opportunity to probe quantum sensing limits and search for physics beyond the standard model.
\end{abstract}

% insert suggested PACS numbers in braces on next line
% insert suggested keywords - APS authors don't need to do this
%\keywords{}

%\maketitle must follow title, authors, abstract, \pacs, and \keywords
\maketitle

Measuring the amplitude of mechanical oscillators has engaged physicists for more than 50 years \citep{Weber1966, Caves1980} and, as the limits of amplitude sensing have dramatically improved, produced exciting advances both in fundamental physics and in applied work. Examples include the detection of gravitational waves \citep{Abbott2016}, the coherent quantum control of mesoscopic objects \citep{Aspelmeyer2014}, improved force microscopy \citep{Butt2005}, and the transduction of quantum signals \citep{Palomaki2013}. During the past decade, optomechanical systems have facilitated increasingly sensitive techniques for reading out the amplitude of a mechanical oscillator \citep{Teufel2009, Anetsberger2010, Westphal2012, Schreppler2014a, Kampel2016}, with a recent demonstration obtaining a measurement imprecision more than two orders of magnitude below $z_{ZPT}$, the amplitude of the ground-state zero-point fluctuations \citep{Wilson2014a}. Optomechanical systems have assumed a wide range of physical systems, including toroidal resonators, nanobeams, and membranes, but the basic principle involves coupling the amplitude of a mechanical oscillator to the resonant frequency of an optical cavity mode \citep{Aspelmeyer2014}.

Crystals of laser-cooled, trapped ions behave as atomic-scale mechanical oscillators \citep{Jost2009,Biercuk2010,Sawyer2012} with tunable oscillator modes and high quality factors ($ {\sim} 10^6$). Furthermore, laser cooling enables ground-state cooling and non-thermal state generation of these oscillators. Trapped-ion crystals therefore provide an ideal experimental platform for investigating the fundamental limits of amplitude sensing. Prior work has demonstrated the detection of coherently driven amplitudes larger than the zero-point fluctuations of the trapped ion oscillator \citep{Biercuk2010,Sawyer2012,Shaniv2016}, and reported impressive force sensing by injection locking an optically amplified oscillation of a single trapped ion \citep{Knunz2010}.

In this Letter we experimentally and theoretically analyze a technique to measure the center-of-mass (COM) motion of a two-dimensional, trapped-ion crystal of $\sim$100 ions with a sensitivity below $z_{ZPT}$. We employ a time-varying spin-dependent force $F_0\cos\left(\mu t\right)$ that couples the amplitude of the COM motion with the internal spin degree of freedom of the ions \citep{Bollinger2013,Sawyer2014,Ivanov2016}. When the frequency $\mu$ matches the frequency $\omega$ of a driven COM oscillation, $Z_{c}\cos\left(\omega t\right)$, spin precession proportional to $Z_{c}$ occurs. The amplitude-dependent spin precession is analogous to the optomechanical frequency shift of a cavity mode. In contrast to the continuous measurement typical of optomechanics experiments, we measure the spin precession only at the end of the experimental sequence, with a precision imposed by spin projection noise \citep{Itano1993}.

To determine the read-out imprecision in a regime free from thermal noise, we perform measurements where $\omega$ is far from resonance with the trap axial frequency $\omega_z$. Additionally, we implement a protocol where the phase of the measured quadrature randomly varies from one iteration of the experiment to the next, appropriate for sensing a force whose phase is unknown or not stable. For $N = 85$ ions and $z_{ZPT} \equiv \frac{1}{\sqrt{N}}\sqrt{\frac{\hbar}{2m\omega_z}} \approx 2\:\mathrm{nm}$, we detect amplitudes $Z_c=500$ pm in a single implementation of the experimental sequence, and as small as 50~pm after averaging over 3,000 iterations of the sequence.

\begin{figure}
    \centering
    \includegraphics[width=\columnwidth]{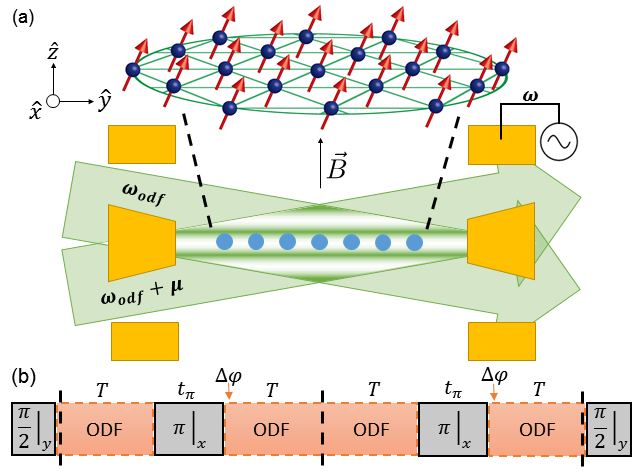}
    \caption{(a) Representation of ion spins arranged in a 2D triangular lattice, along with a cross-sectional illustration of the Penning trap, characterized by an axial magnetic field $B = 4.45$ T and an axial trap frequency $\omega_z = 2\pi \times 1.57$ MHz. The blue dots represent ions. Cylindrical electrodes (yellow) generate a harmonic confining potential along the $\hat{z}$-axis. Radial confinement is provided by the Lorentz force from $\vec{E} \times \vec{B}$-induced rotation in the axial magnetic field. The beams generating the spin-dependent optical-dipole force (green arrows) cross the ion plane at $\pm 10^{\circ}$, forming a 1D traveling-wave potential (green lines) with $\delta k = 2 \pi /(0.9 \, \mu \rm{m})$. An AC voltage source is connected to the trap endcap and used to drive an axial oscillation with calibrated amplitude $Z_c$. (b) Quantum lock-in CPMG sequence used to detect spin precession produced by COM motion resonant with the ODF. Doppler cooling and $\ket{\uparrow}_N$ spin-state preparation occur before the sequence, and spin-state detection after. Grey blocks with solid borders represent microwave $\pi/2$ rotations about $\hat{y}$ and $\pi$ rotations about $\hat{x}$. Orange blocks with dashed borders represent ODF pulses. The ODF phase is advanced by $\Delta\varphi$ in a modulation scheme discussed in \citep{SuppMat}, where $\Delta\varphi = \pi$ for $\omega = \mu$. Dashed vertical lines indicate the $m$ segments of the sequence, here $m = 2$. We make use of an $m = 8$ sequence for Figs. 2-4.} 
    \label{Expt}
\end{figure}

Our experimental apparatus, described in Fig. \ref{Expt} and \citep{Bollinger2013,Sawyer2014,Bohnet2015}, consists of $N\sim100$ $\prescript{9}{}{}$Be$^{+}$ ions laser-cooled to the Doppler limit of 0.5 mK and confined to a single-plane Coulomb crystal in a Penning trap. The spin-1/2 degree of freedom is the $\prescript{2}{}{S}_{1/2}$ ground-state valence electron spin $\ket{\uparrow} (\ket{\downarrow}) \equiv \ket{m_{s}=+1/2} (\ket{m_{s}=-1/2}) $. In the magnetic field of the Penning trap, the spin-flip frequency is 124 GHz. A resonant microwave source is used to perform global rotations of the spin ensemble. A pair of laser beams, detuned from the nearest optical transitions by $\sim$20 GHz, interfere to form a one-dimensional (1D) traveling-wave potential that produces a spin-dependent optical-dipole force (ODF). Optical pumping prepares the initial state $\ket{\uparrow}_N \equiv \ket{\uparrow \uparrow \cdots \uparrow}$ with high fidelity. At the end of the experiment we measure the probability $P_\uparrow$ for an ion spin to be in $\ket{\uparrow}$ from a global measurement of state-dependent resonance fluorescence on the Doppler cooling transition, where spin $\ket{\uparrow}$ ($\ket{\downarrow}$) is bright (dark).

If the ions are localized axially over an extent small compared with the wavelength of the 1D traveling-wave potential (Lamb-Dicke confinement), then the ODF couples the spin and motional degrees of freedom through the interaction \citep{Bohnet2015}
\begin{equation}
\hat{H}_{ODF} = F_0\cos\left(\mu t \right)\sum_{i} \hat{z}_{i} \hat{\sigma}^{z}_{i}.
\label{Hodf}
\end{equation}
Here $F_{0}= U \, \delta k \, \rm{\it{DWF}}$ is the magnitude of the ODF, where $U\:(\delta k)$ is the zero-to-peak potential (wave vector) of the 1D traveling-wave, $\mu$ is the frequency difference between the ODF beams, and $\hat{z}_{i}$ and $\hat{\sigma}^{z}_{i}$ are the position operator and Pauli spin matrix for ion $i$. The Debye-Waller factor $\rm{\it{DWF}} = \exp(-\delta k^2 \left< \hat{z}^{2}_{i} \right> / 2)$ reduces $F_{0}$ due to the departure from the Lamb-Dicke confinement regime \citep{Wineland1998a}; $\rm{\it{DWF}} \approx 0.86 $ for the conditions of this work. The potential $U$, and therefore $F_0$, is determined from AC Stark shift measurements on the ions \citep{Britton2012}. Typical maximum values for this work are $U/\hbar \approx 2 \pi \times (10.4$~kHz) resulting in $F_0 \approx 40$ yN.

Equation (\ref{Hodf}) describes a dependence of the spin transition frequency on the axial position of the ions and the ODF frequency $\mu$. We excite a small, classically driven COM motion of constant amplitude $\hat{z}_i \rightarrow \hat{z}_i +Z_c\cos(\omega t+\delta)$ with a weak RF drive on a trap endcap electrode (see Fig. \ref{Expt}(a)) at a frequency $\omega$ far from $\omega_{z}$. If $\omega\sim\mu$, Eq. (\ref{Hodf}) produces an approximately constant shift in the spin transition frequency. With $\delta k Z_c \ll 1$, this shift is given by
\begin{equation}
\hat{H}_{ODF} \approx F_{0} \, Z_c\cos((\omega - \mu)t + \delta) \sum_{i} \frac{\hat{\sigma}^{z}_{i}}{2} .
\end{equation}
For $\mu = \omega$, the static shift of the spin transition frequency is simply $\Delta(Z_c) = (F_{0}/\hbar) \, Z_c \cos(\delta)$.

We measure $\Delta(Z_c)$ from the resulting spin precession in an experiment like that shown in Fig. \ref{Expt}(b). Ideally, spin precession can be measured using a Ramsey-type experiment. First, the ions are prepared in the $\ket{\uparrow}_N$ state, followed by a microwave $\pi/2$ pulse about $\hat{y}$ that rotates the spins to the $\hat{x}$ axis. The spins precess for an interaction time $\tau$ so that the resulting spin precession on resonance $(\mu = \omega)$ is $\theta = \theta_{max} \cos(\delta)$, where $\theta_{max} \equiv (F_{0}/\hbar)\, Z_c \, \tau$. After a final $\pi/2$ pulse about $\hat{y}$, the final state readout measures the population of the spins in $\ket{\uparrow}$, $P_{\uparrow} = \frac{1}{2}[1-e^{-\Gamma \tau}\cos(\theta)]$. Here $\Gamma$ is the decay rate from spontaneous emission from the off-resonant ODF laser beams \citep{Uys2010}. To detect small amplitudes with the available $F_0$ in our set-up, we extend the spin-precession time to $\tau \ge 20$ ms. To avoid decoherence due to magnetic field fluctuations and coherently accumulate spin precession, we use a quantum lock-in \citep{Kotler2011} sequence where during the interaction time $\tau$ the spin precession is interrupted by a train of $\pi$-pulses that are synchronized with phase jumps enforced on the ODF beams \citep{SuppMat}. In particular, we use a Carr-Purcell-Meiboom-Gill (CPMG) sequence with $m = 8$ \mbox{ODF-$\pi$-ODF segments} ($\tau = 2\, m \, T$, see Fig. \ref{Expt}(b)).

\begin{figure}
    \centering
    \includegraphics[width=\columnwidth]{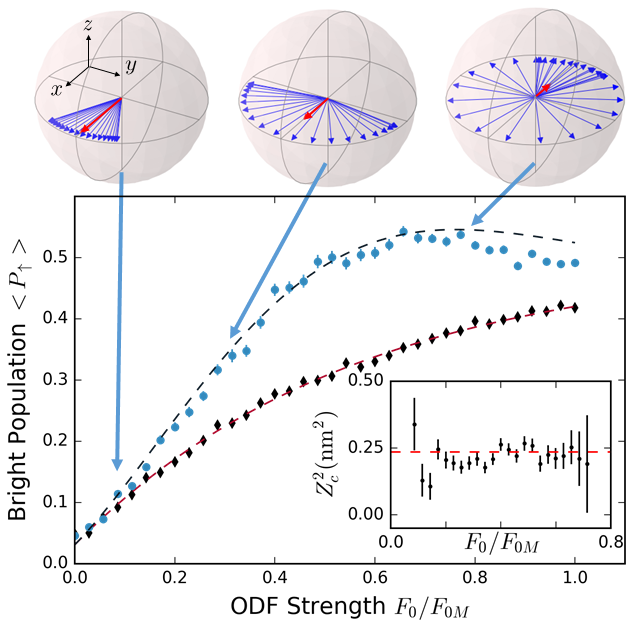}
  \caption{\textbf{Top:} Bloch sphere representation \citep{QuTip} of spin dephasing for $Z_{c} = 485\: \rm{pm}$. Each blue vector represents an experimental trial with a different phase $\delta$ (see text). From left to right, the spread in the blue vectors corresponds to $\theta_{max} = 0.470, 1.41, 3.62$ radians and $F_{0}/F_{0M} = 0.1,0.3,0.77$, where $F_{0M}$ is the maximum optical-dipole force. Our experiment measures the length of the Bloch vector averaged over many trials, denoted by the thick red vector. \textbf{Main plot:} As a function of ODF strength, the background (black diamonds) with no applied drive and signal (blue points) for a 485 pm amplitude and total ODF interaction time $\tau$ = 24 ms is shown. The red dashed line is a fit to the background. The black dashed line is the prediction with no free parameters, given the background fit. Here $N = 75$ ions and $F_{0M} = 41.3$ yN. \textbf{Inset:} Black points are experimentally determined values for $Z_{c}^2$. Red dashed line is the calibrated value of $Z_{c}^{2}$. Error bars represent standard error.}\label{Meas_stren}
\end{figure}

We ensure the phase $\delta$ randomly varies from one iteration of the CPMG sequence to the next, effectively measuring a random quadrature of the motion for each experimental trial. Different experimental trials therefore result in a different precession $\theta$, as indicated in Fig. \ref{Meas_stren}. We measure the collective dephasing (or decoherence) averaged over many experimental trials $\left< P_{\uparrow} \right> = \frac{1}{2}[1-e^{-\Gamma \tau} \left<\cos(\theta)\right>]$. Here the brackets $ \left< \cdot \right> $ denote an average over many iterations of the CPMG sequence. Averaging over the random phase $\delta$ yields~\citep{Kotler2013}
\begin{equation}
\left< P_{\uparrow} \right> = \frac{1}{2} \left[ 1-e^{-\Gamma \tau}J_0(\theta_{max}) \right],
\label{Bessel}
\end{equation}
with $J_0$ the zeroth-order Bessel function of the first kind.

To create the steady-state COM axial oscillation $Z_c \cos(\omega t+\delta)$, we applied a continuous AC voltage to an endcap of the Penning trap at a frequency $\omega/(2\pi)$ near 400 kHz. This frequency was chosen because it was far from any motional mode frequencies of the ion crystal, and there were no observed noise sources. Thus, the background, i.e. the signal without the driven COM axial motion such that $Z_c =0$, was fully characterized by decoherence due to spontaneous emission and is given by $\left\langle P_{\uparrow}\right\rangle _{bck}= \frac{1}{2}\left[1-e^{-\Gamma\tau}\right]$. We calibrated the displacement of the ions due to a static voltage applied to the endcap by measuring the resulting movement of the ion crystal in the side-view imaging system. From this calibration, we determined that a 1 V offset results in a 0.97(5) $\mu$m displacement of the ions. We estimate that the corrections for using this DC calibration to estimate $Z_c$ for an $\omega/(2\pi) \approx 400$ kHz drive is less than 10$\,\%$.

\begin{figure}
    \centering
    \includegraphics[width=\columnwidth]{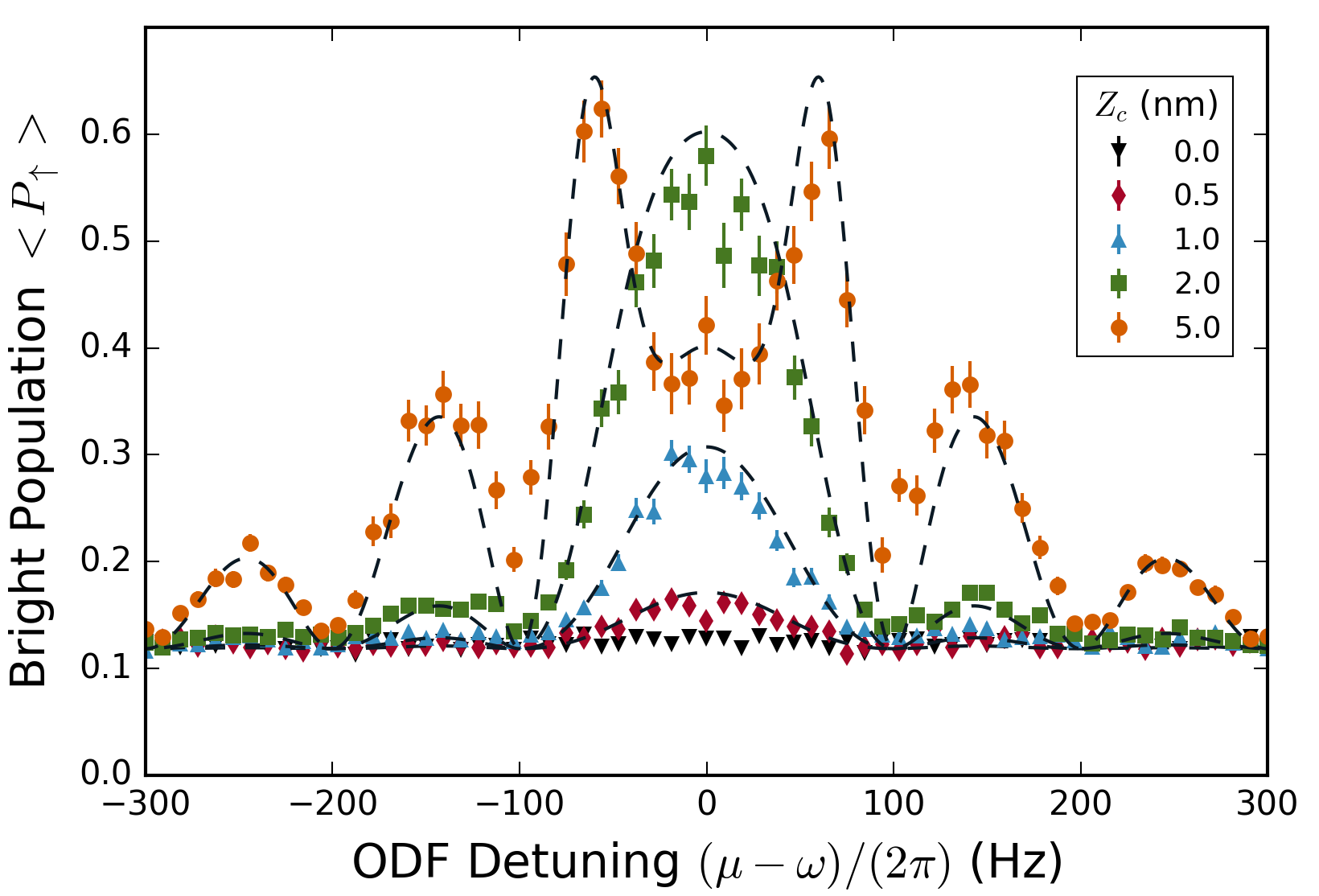}
  \caption{Lineshape of the spin precession signal for amplitudes $Z_c$ of 500 pm (red diamonds), 1 nm (blue triangles), 2 nm (green squares), and 5 nm (orange circles) for $\tau$ = 20 ms. Black triangles are the background, with the drive turned off. Dashed lines are predictions with no free parameters. Error bars represent standard error. Here $N = 90$ ions and $F_{0} = 7.9$ yN.}\label{lineshape}
\end{figure}

Figure \ref{lineshape} shows the emergence of the measured spin precession signal out of the background as the amplitude $Z_c$ is increased from 500 pm to 5 nm. The measured lineshape agrees well with the prediction, detailed in \citep{SuppMat}, involving no free parameters. Figure \ref{Meas_stren} shows the background and the measured resonant ($\mu=\omega$) response to a $Z_c$ = 485 pm oscillation for a range of ODF strengths $F_{0}/F_{0M}$, where $F_{0M}$ is the maximum $F_0$ possible with our current set-up ($\sim 40$ yN). Agreement with Eq. (\ref{Bessel}) involving no free parameters is excellent. For both Figs. \ref{lineshape} and \ref{Meas_stren} the background is within $6\,\%$ of that determined by independent measurements of the spontaneous emission decay rates of each ODF beam \citep{Britton2012}. The amplitude $Z_c=\theta_{max}/(\tau F_{0}/\hbar)$ can be determined from the difference $\left< P_\uparrow \right>- \left< P_\uparrow \right>_{bck}$ \citep{SuppMat}. We note that $\left< P_\uparrow \right>- \left< P_\uparrow \right>_{bck}$ depends on $\theta_{max}^2$. Therefore, the sensing protocol described here directly measures $Z_c^2$. The inset of Fig. \ref{Meas_stren} shows a determination of $Z_c^2$ for a range of ODF strengths. The uncertainties were calculated from the measured noise of the $\left< P_\uparrow \right>- \left< P_\uparrow \right>_{bck}$ measurements using standard error propagation. These uncertainties go through a minimum, indicating an optimum $F_{0}/F_{0M}$ value for determining $Z_c^2$.

\begin{figure}
\includegraphics[width=\columnwidth]{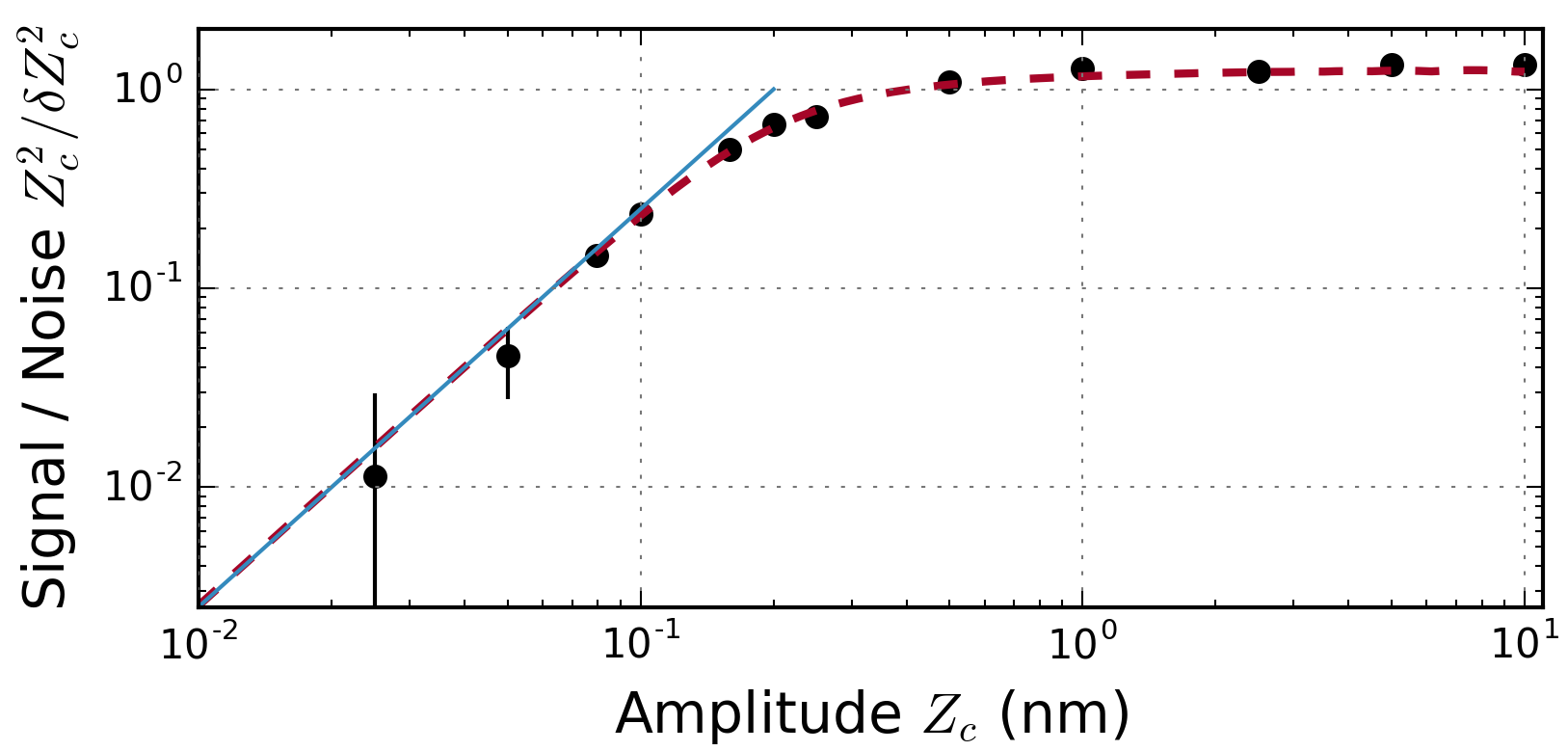}
\caption{Amplitude sensing limits for $N=85$. Black points are the experimentally measured signal-to-noise for determinations of $Z_{c}^{2}$ from single pairs of $\ P_{\uparrow},\: P_{\uparrow, bck}$ measurements as a function of the experimentally imposed $Z_c$. Our measurement for $Z_c =$ 25 pm is consistent with zero. Red dashed line is the prediction for the signal-to-noise including projection noise and the random COM mode quadrature measured each trial. Blue solid line is the predicted limiting signal-to-noise for small amplitudes (Eq. (\ref{eq:limiting})), assuming only projection noise and parameters relevant for our set-up. Error bars represent standard error.} \label{Fig_sens}
\end{figure}
To explore the ultimate amplitude sensing limits of our protocol, we performed repeated pairs of $P_{\uparrow}$ measurements, first with $Z_c = 0$ to get the background, and then with \mbox{$Z_{c}\neq0$}. For a given $Z_{c}$, 3,000 pairs of measurements were used to determine the average difference $\left\langle P_{\uparrow}\right\rangle -\left\langle P_{\uparrow}\right\rangle _{bck}$ and the standard deviation $\sigma\left( P_{\uparrow} - P_{\uparrow ,bck} \right)$ of the difference for a single pair of measurements. For each $Z_c$, $F_{0}/F_{0M}$ was set close to the value that maximizes the signal-to-noise ratio for determining $Z_{c}^{2}$. This occurs for relatively small $\theta_{max}$ such that \mbox{$\frac{1}{2}\left(1-J_{0}\left(\theta_{max}\right)\right)\approx\theta_{max}^{2}/8$}. Then, the signal-to-noise ratio for determining $Z_{c}^{2}$ from a single pair of $\ P_{\uparrow},\: P_{\uparrow, bck}$ measurements is approximately 
\begin{equation}
\frac{Z_{c}^{2}}{\delta Z_{c}^{2}}\approx\frac{\left\langle P_{\uparrow}\right\rangle -\left\langle P_{\uparrow}\right\rangle  _{bck}}{\sigma\left( P_{\uparrow} - P_{\uparrow , bck} \right)}\:.\label{eq:signal/noise}
\end{equation}
Figure \ref{Fig_sens} displays Eq. (\ref{eq:signal/noise}) from measurements acquired with $Z_{c}$
ranging from 10 nm to as small as 0.025 nm. Excellent agreement is observed with a model (dashed red line) that assumes the only noise sources are projection noise in the spin-state detection and fluctuations in $P_{\uparrow}$ produced by random variation in the phase $\delta$ from one experimental trial to the next.

For amplitudes $Z_{c}\gtrsim 500\:\mathrm{pm}$, fluctuations in $P_{\uparrow}$ due to the random variation of the phase $\delta$ for different experimental trials dominates. This situation is depicted by the middle Bloch sphere of Fig. \ref{Meas_stren}. Here the fluctuations in $P_{\uparrow}$ are comparable to the difference $\langle P_{\uparrow}\rangle - \langle P_{\uparrow} \rangle_{bck}$, limiting the signal-to-noise of a single determination of $Z_{c}^{2}$ to ${\sim} 1$. As $Z_{c}$ decreases, this noise and the signal decrease while projection noise stays approximately the same, resulting
in a decreasing $Z_{c}^{2}/\delta Z_{c}^{2}$. For small $Z_c$, we show the sensitivity is determined by $N$, $\delta k$, and the ratio of the spontaneous decay rate to the optical potential $\xi\equiv\Gamma/\left(U/\hbar\right)$ \citep{SuppMat}, according to
\begin{equation}
\left.\frac{Z_{c}^{2}}{\delta Z_{c}^{2}}\right|_{\mathrm{limiting}}\approx 0.097 \frac{\sqrt{N}(\rm{\it{DWF}})^2 (\delta k)^2}{\xi^2} Z_c^2\:.\label{eq:limiting}
\end{equation}
For $N=85$ and values of $\rm{\it{DWF}}$, $\delta k$, and $\xi=1.156\times10^{-3}$ relevant for our set-up, Eq. (5) predicts $Z_c^2/\delta Z_c^2 \approx \left[ Z_c/ 0.2 \, \rm{nm} \right]^2$, displayed as the blue line in Fig. \ref{Fig_sens}. On the log-log plot the slope of 2 is the result of a signal proportional to $Z_c^2$ along with a constant readout noise of the spins (here projection noise). We perform 16 pairs of measurements in 1 s, so the signal-to-noise ${Z_{c}^{2}}/{\delta Z_{c}^{2}} \approx \left[ Z_c/ 0.2 \, \rm{nm} \right]^2$ for a single pair of measurements corresponds to a long averaging time sensitivity of $\left(100\:\mathrm{pm}\right)^{2}/\sqrt{\mathrm{Hz}}$ (recall that our protocol measures $Z_{c}^2$).

Figure \ref{Fig_sens} documents a good understanding of the sensing limits of our protocol, indicating how the measurement can be improved in the future. Equation (\ref{eq:limiting}) scales as $1/\xi^{2}$, resulting in significant improvements for set-ups with less spontaneous decay. By stabilizing the ODF beatnote phase with respect to the classical drive \citep{Hume2011, Biercuk2011} we could repeatedly measure the same quadrature of motion and realize a substantial improvement in sensitivity. For this phase-coherent protocol, assuming $N=100$ and current parameters of our set-up, we estimate \citep{SuppMat} a measurement imprecision of 74 pm for a single implementation of the experimental sequence. This is ${\sim} \,30$ times smaller than $z_{ZPT}$, producing a long averaging time sensitivity of ${\sim}18\:\mathrm{pm/\sqrt{\mathrm{Hz}}}$. The use of spin-squeezed states, recently demonstrated in this system \citep{Bohnet2015}, can provide an additional enhancement by reducing the projection noise of the readout.

The 50 pm amplitude detected in Fig. \ref{Fig_sens} at a frequency $\omega$ far from resonance corresponds to an electric field detection of 0.46 mV/m or 73 yN/ion. These force and electric field sensitivities can be improved by the $Q$ of the COM mode by probing near resonance with $\omega_z$. Quality factors $Q\sim 10^6$ should be possible with trapped-ion COM modes. The detection of a 20 pm amplitude resulting from a 100 ms coherent drive on the 1.57 MHz COM mode is sensitive to a force/ion of $5\times10^{-5}\:\mathrm{yN}$ corresponding to an electric field of $0.35\:\mathrm{nV/m}$. Electric field sensing below ${\sim} \,1$ nV/m enables searches for hidden-photon dark matter \citep{Arias2012,Chaudhuri2015}, although shielding effects must be carefully considered. Ion traps typically operate with frequencies $\omega_z/2\pi$ between 50 kHz and 5 MHz, providing a sensitivity to hidden-photon masses from $2 \times 10^{-10}$ eV to $2 \times 10^{-8}$ eV. 

By sensing COM motion far from resonance, we calibrate the measurement imprecision of our protocol in the absence of thermal noise and back action. Probing on resonance with a measurement imprecision below $z_{ZPT}$ will be sensitive to thermal fluctuations and back action due to spin-motion entanglement \citep{Sawyer2014}. This motivates the investigation of potential back-action-evading protocols with trapped ion set-ups. For the phase coherent measurement of a single quadrature, back action due to spin-motion entanglement can be evaded through the introduction of the appropriate correlations between spin and motion \citep{Hempel2013}.

In summary, we have presented a technique for amplitude sensing below $z_{ZPT}$ of a trapped ion mechanical oscillator. By employing a spin-dependent force to couple the spin and motional degrees of freedom of the ions, the amplitude of the COM motion may be determined. We detected a 500 pm amplitude in a single experimental trial and demonstrated a long measurement time sensitivity of $\left(100\:\mathrm{pm}\right)^{2}/\sqrt{\mathrm{Hz}}$ with a protocol where the phase of the measured quadrature randomly varies. Modifications of our set-up should enable repeated measurements of the same quadrature, with a measurement imprecision of 74 pm for a single experimental trial with $N = 100$ ions, providing opportunities for trapped ion mechanical oscillators to explore the quantum limits of amplitude and force sensing, and enable new tools in the search for physics beyond the standard model.

\begin{acknowledgments}
We thank V. Sudhir, R. Ozeri, S. Kotler, J. Teufel, J. Jaeckel, J. E. Jordan, and D. Kienzler for stimulating discussions. K.G. is supported by NSF grant PHY 1521080. This manuscript is a contribution of NIST and not subject to U.S. copyright.
\end{acknowledgments}

\bibliographystyle{apsrev4-1}
\bibliography{amplitude_sensing}

\end{document}

\title{Supplemental Material: Amplitude sensing below the zero-point fluctuations with a two-dimensional trapped-ion mechanical oscillator}
\author{K. A. Gilmore}
\email[]{kevin.gilmore@colorado.edu}

\affiliation{National Institute of Standards and Technology, Boulder, Colorado 80305, USA}
\affiliation{JILA and Department of Physics, University of Colorado, Boulder, Colorado, 80309, USA}

\author{J. G. Bohnet}
\affiliation{National Institute of Standards and Technology, Boulder, Colorado 80305, USA}

\author{B. C. Sawyer}
\affiliation{Georgia Tech Research Institute, Atlanta, Georgia 30332, USA}

\author{J. W. Britton}
\affiliation{U.S. Army Research Laboratory, Adelphi, Maryland 20783, USA}

\author{J. J. Bollinger}
\email[]{john.bollinger@nist.gov}
\affiliation{National Institute of Standards and Technology, Boulder, Colorado 80305, USA}
\maketitle

\section*{Introduction}
In this supplemental material we provide detailed derivations for a number of theoretical formulas and results of the main text. Specifically, in the first section we derive  the shift in the spin transition frequency due to the coherent amplitude $Z_c$ (Eq. (2) of the main text) from a more basic perspective. We also discuss in detail our modulation scheme. In the second section we derive the lineshape function used in Fig. 2 of the main text. In section 3 we describe the formalism used to determine the optimum signal-to-noise ratio for a measurement of $Z_{c}^2$. We used this optimum signal-to-noise ratio to generate the theoretical curve in Fig. 4 of the main text. We also derive the sensitivity limits for phase-incoherent amplitude sensing, where the phase difference between the driven motion and the ODF randomly varies from one realization of the experiment to the next. We show how these limits depend on $\Gamma/(U/\hbar)$, $\delta k$, and $N$. Finally, in section 4 we consider the amplitude sensing limits assuming phase coherence between the spin-dependent force and the driven amplitude. 

\tableofcontents

\section{1. Shift in the spin transition frequency, and the modulation scheme}
Figure 1 shows the Carr-Purcell-Meiboom-Gill (CPMG) sequence used to apply the 1D traveling-wave potential. The interaction of the spin degree of freedom with the 1D traveling-wave potential is given by
\begin{equation}
\hat{H}_{ODF} = U\sum_{i}\sin(\delta k \cdot \hat{z}_{i} - \mu t + \phi)\hat{\sigma}^{z}_{i} = U\sum_{i}\sin(\delta k \cdot \hat{z}_{i})\cos(\mu t - \phi)\hat{\sigma}^{z}_{i} - U\sum_{i}\cos(\delta k \cdot \hat{z}_{i})\sin(\mu t - \phi)\hat{\sigma}^{z}_{i}.
\label{}
\end{equation}
Here we explicitly include a phase $\phi$ for the traveling-wave potential. Without loss of generality, we assumed $\phi = 0$ in the main text. If $\delta k \left< \hat{z}_{i} \right> \ll 1$, then $\left< \cos(\delta k \cdot \hat{z}_{i}) \right> \sim 1$, and the spin precession due to the second term will be bounded by $(U/\hbar)/\mu$.

Typically, $(U/\hbar)/\mu \ll 1$ and thus this term is ignored in most treatments. At low frequencies $\mu \le U/\hbar$ this term could be important, but it may be canceled by advancing the phase of the ODF by $\Delta \phi = \mu(T+t_{\pi})$ at each microwave $\pi-$pulse of the CPMG sequence (see Fig. 1). When $\mu/2\pi = (2n+1)/(2(T+t_{\pi}))$ for some integer $n$, $\Delta\varphi = \pi$ and we recover the quantum lock-in phase advance of \citep{Kotler2011}. This phase advance coherently accumulates spin precession from the first term of Eq. (1) when $\omega/2\pi = (2n+1)/(2(T+t_{\pi}))$. The term that survives our modulation scheme is
\begin{equation}
\hat{H}_{ODF} \simeq U\sum_{i}\sin(\delta k \cdot \hat{z}_{i})\cos(\mu t  - \phi)\hat{\sigma}^{z}_{i}.
\label{}
\end{equation}

We now impose a weak, classically driven COM motion of constant amplitude and phase $\hat{z}_i \rightarrow \hat{z}_i +Z_c\cos(\omega t+\delta)$. This can be thought of as the center of the Penning trap being moved by $\pm Z_c$ at a frequency $\omega$ far from the trap axial frequency $\omega_z$. With $\delta k \, Z_c \ll 1$, we obtain
\begin{equation}
\hat{H}_{ODF} \simeq  U\sum_{i} \left( \delta k \, Z_c \cos(\delta k \cdot \hat{z}_{i})\cos(\omega t + \delta)\cos(\mu t - \phi) + \sin(\delta k \cdot \hat{z}_{i})\cos(\mu t - \phi) \right) \hat{\sigma}^{z}_{i}.
\label{}
\end{equation}
The second term of Eq. (3) is the usual term that gives rise to spin-motion entanglement with the drumhead modes and to effective spin-spin interactions \citep{[][{ (see Supplementary Information).}]Britton2012,[][{ (see Supplementary Information).}]Bohnet2015}. We assume we can neglect this term because we tune $\mu$ far from any drumhead modes.

Deep in the Lamb-Dicke confimenent regime, the $\cos(\delta k \cdot \hat{z}_{i})$ factor in the first term of Eq. (3) equals one. Here we account for the possibility of not being deep in the Lamb-Dicke confinement regime. In this case, and assuming a thermal distribution of modes, $\left< \cos(\delta k \cdot \hat{z}_{i}) \right> = \exp(-\delta k^2 \left< \hat{z}^{2}_{i} \right> / 2) $. This factor is known as the Debye-Waller factor $\rm{\it{DWF}}$. For our conditions all ions have approximately the same Debye-Waller factor, $\rm{\it{DWF}} \approx 0.86$ \citep{Bohnet2015}.

With $\mu \sim \omega$, Eq. (3) can be written as
\begin{equation}
\hat{H}_{ODF} = (U \cdot \delta k \cdot \rm{\it{DWF}}) \,  Z_c\cos((\omega - \mu)t + \delta - \phi) \sum_{i} \frac{\hat{\sigma}^{z}_{i}}{2},
\label{}
\end{equation}
which is Eq. (2) of the main text with $F_0 = U \cdot \delta k \cdot \rm{\it{DWF}}$.

\section{2. Lineshape}
To model the lineshape of the signal, it is necessary to account for the accumulated phase due to the spin-dependent ODF potential without making the simplification that $ \omega = \mu $. This results in a characteristic response function for each sequence. For this Letter, we used an $ m = 8 $ CPMG sequence, as shown in Fig. 1 in the main text and also Fig. 1 of this supplemental material. In the following, we derive the lineshape of this sequence using the modulation discussed in the first section and assuming a delta function source at a frequency $\omega$. This lineshape is used to generate the theory curves of Fig. 2 of the main text. In general, for a CPMG sequence it is necessary to calculate the phase evolution during $2m$ terms of length T, for a total interaction time of $2m$T. For simplicity, we first derive the lineshape for the $ m = 2 $ CPMG sequence (Fig. 1).
\begin{figure}
    \centering
    \includegraphics[width=1.0\textwidth]{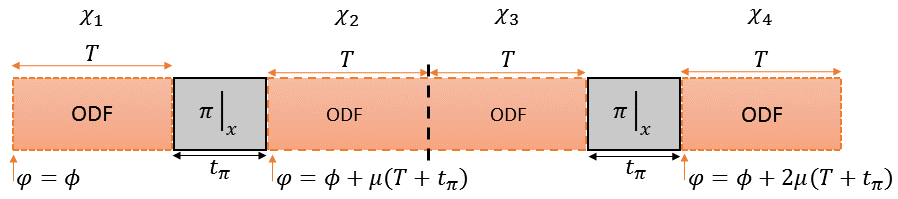}
    \caption{$m = 2$ CPMG sequence with total ODF interaction time $4T$. $\varphi$ is the phase of the ODF beatnote. The $\chi_i$ labels represent the periods over which the accumulated phase is considered in the text.}
    \label{cpmg_sup}
\end{figure}

 For a delta function source $Z_c \cos(\omega t+\delta)$, the spin precession accumulated in a general sequence like that shown in Fig. 1 is

\begin{equation}
\theta(\mu) = F_0 \, Z_c \frac{2 \sin \left( \frac{1}{2}\left(\omega-\mu \right)T \right)}{(\omega-\mu)} \chi(\mu,\omega),
\label{}
\end{equation}
where $\chi(\mu,\omega) = \sum_{i} \chi_i(\mu,\omega)$ is determined by the particular sequence used. In the case of the $ m = 2 $ CPMG sequence, the phase accumulated through four terms corresponding to four separate applications of the ODF (Fig. 1) must be considered:

\begin{equation}
\chi_{1} = \cos \left[  \left( \omega - \mu \right) \frac{T}{2}  + \delta-\phi \right],
\end{equation}

\begin{equation}
\chi_{2} = -\cos \left[  \left( \omega - \mu \right) \left( \frac{3T}{2} + t_{\pi} \right) +\delta-\phi +\mu(T+t_{\pi}) \right],
\end{equation}

\begin{equation}
\chi_{3} = -\cos \left[  \left( \omega - \mu \right) \left( \frac{5T}{2} + t_{\pi} \right) +\delta-\phi +\mu(T+t_{\pi}) \right],
\end{equation}

\begin{equation}
\chi_{4} = \cos \left[  \left( \omega - \mu \right) \left( \frac{7T}{2} + 2t_{\pi} \right) +\delta-\phi +2\mu(T+t_{\pi}) \right].
\end{equation}

Note these terms now include a phase $\phi$ for the ODF interaction, which in the main text we set to zero with no loss of generality. Adding these terms up, pairwise:

\begin{equation}
\chi_{1} + \chi_{2} = 2 \sin \left( \frac{1}{2} \left[  \left( \omega - \mu \right) \left( T + t_{\pi} \right) +\mu(T+t_{\pi}) \right] \right) \sin \left[  \left( \omega - \mu \right) \left( T + \frac{t_{\pi}}{2} \right) +\delta-\phi + \frac{\mu(T+t_{\pi})}{2} \right],
\end{equation}

\begin{equation}
\chi_{3} + \chi_{4} = -2 \sin \left( \frac{1}{2} \left[  \left( \omega - \mu \right) \left( T + t_{\pi} \right) +\mu(T+t_{\pi}) \right] \right) \sin \left[  \left( \omega - \mu \right) \left( 3T + \frac{3t_{\pi}}{2} \right) +\delta-\phi + \frac{3\mu(T+t_{\pi})}{2} \right].
\end{equation}

Summing all four terms yields

\begin{equation}
\chi(\mu,\omega) = \sum_{i} \chi_i(\mu,\omega) = 2 \sin \left( \frac{\omega}{2}\left( T + t_{\pi} \right) \right) \left[ \sin \left( \xi + \delta-\phi \right) - \sin \left( 3\xi + \delta-\phi \right) \right],
\end{equation}
where $\xi = (\omega-\mu)(T+\frac{t_{\pi}}{2})+\frac{\mu(T+t_{\pi})}{2} = \frac{1}{2}\left( \omega (T+t_{\pi}) + T \left( \omega-\mu \right) \right)$.
Then, simplifying:
\begin{equation}
\chi(\mu,\omega) = 2 \sin \left( \frac{\omega}{2}\left( T + t_{\pi} \right) \right) 2 \sin \left(- \xi \right) \cos \left( 2 \xi + \delta-\phi \right).
\end{equation}

Using Eqs. 13 and 5,
\begin{equation}
\theta(\mu) = \rm{\it{DWF}} \cdot U \cdot \delta k \cdot Z_c \cdot T \sinc \left( \frac{T}{2}\left(\omega-\mu \right) \right) 4 \sin \left( \frac{\omega}{2}\left( T + t_{\pi} \right) \right) \sin \left( \xi \right) \cos \left( 2 \xi + \delta-\phi \right).
\end{equation}

Since $4T = \tau$ for the $m = 2$ CPMG, then

\begin{equation}
\theta(\mu) = \theta_{max} \sinc \left( \frac{T}{2}\left(\omega-\mu \right) \right) \sin \left( \frac{\omega}{2}\left( T + t_{\pi} \right) \right) \sin \left( \xi \right) \cos \left( 2 \xi + \delta-\phi \right),
\end{equation}
where $\theta_{max} \equiv (F_{0}/\hbar)\, Z_c \, \tau$, the maximum precession angle on resonance as defined in the main text. Then, $\theta_{max}(\mu)$, defined as $\theta(\mu) = \theta_{max}(\mu) \cos \left( 2 \xi + \delta-\phi \right)$, is the $\mu$-dependent generalization of $\theta_{max}$. From Eq. 15, this is
\begin{equation}
\theta_{max}(\mu) = \theta_{max} \sinc \left( \frac{T}{2}\left(\omega-\mu \right) \right) \sin \left( \frac{\omega}{2}\left( T + t_{\pi} \right) \right) \sin \left( \xi \right).
\end{equation}
For the $m = 8$ CPMG sequence the same procedure is used, but now with 16 periods of accumulated phase. We obtain

\begin{widetext}
\begin{equation}
\theta_{max}(\mu) = \theta_{max}  \sinc\left( \frac{T}{2} \left( \omega-\mu \right) \right) \sin\left( \frac{\omega}{2} \left( T+t_{\pi} \right) \right) \sin(\xi)\cos(2 \xi)\cos(4 \xi).
\end{equation}
\end{widetext} 

As shown in the main text, the expression for population in $\ket{\uparrow}$ - now with a dependence on the ODF difference frequency $\mu$ - is

\begin{equation}
\left< P_{\uparrow} \right> = \frac{1}{2} \left[ 1-e^{-\Gamma \tau}J_0(\theta_{max}(\mu)) \right].
\label{Bessel}
\end{equation}
Equations (17) and (18) are used to obtain the theoretical line shapes of Fig. (2) of the main text.
\section{3. Phase-incoherent sensing limits}

Here we derive Eq. (5) from the main text and provide additional mathematical background for the phase-incoherent experimental protocol, wherein the phase of the measured quadrature varies randomly from one realization of the experiment to the next. Following earlier discussions, the probability of measuring $\left|\uparrow\right\rangle $
at the end of the Ramsey sequence is 
\begin{equation}
\left\langle P_{\uparrow}\right\rangle =\frac{1}{2}\left[1-e^{-\Gamma\tau}J_{0}\left(\theta_{max}\right)\right]\:,\label{eq:P_up formula}
\end{equation}
where $\left\langle \:\right\rangle $ denotes an average over many
experimental trials and therefore over the random phase between the
1D traveling-wave potential and the classically driven COM motion, and
\begin{equation}
\theta_{max}=(F_{0}/\hbar) \cdot Z_{c}\cdot\tau\:.\label{eq:theta_max formula}
\end{equation}
Defining $G\left(\theta_{max}^{2}\right)\equiv\left(1-J_{0}\left(\theta_{max}\right)\right)/2$
and denoting $\left\langle P_{\uparrow}\right\rangle _{bck}=\left[1-e^{-\Gamma\tau}\right]/2$
as the probability of measuring $\left|\uparrow\right\rangle $ at
the end of the sequence in the absence of a classically driven motion,
$\theta_{max}^{2}$ can be determined from a measurement of the difference
$\left\langle P_{\uparrow}\right\rangle -\left\langle P_{\uparrow}\right\rangle _{bck}$
through
\begin{equation}
G\left(\theta_{max}^{2}\right)=e^{\Gamma\tau}\left(\left\langle P_{\uparrow}\right\rangle -\left\langle P_{\uparrow}\right\rangle _{bck}\right)\:.\label{eq:P_up difference}
\end{equation}
The standard deviation $\delta\theta_{max}^{2}$ in estimating $\theta_{max}^{2}$
is determined from the standard deviation $\sigma\left(P_{\uparrow} - P_{\uparrow, bck}\right)$
of the $\left\langle P_{\uparrow}\right\rangle -\left\langle P_{\uparrow}\right\rangle _{bck}$
difference measurements through
\begin{equation}
\delta\theta_{max}^{2}=\frac{e^{\Gamma\tau}\sigma\left(\left\langle P_{\uparrow}\right\rangle -\left\langle P_{\uparrow}\right\rangle _{bck}\right)}{\frac{\mathrm{d}G\left(\theta_{max}^{2}\right)}{d\theta_{max}^{2}}}\:.\label{eq:delta theta^2}
\end{equation}
The signal-to-noise ratio of a measurement of $\theta_{max}^{2}$
(and therefore $Z_{c}^{2}$) is $\theta_{max}^{2}/\delta\theta_{max}^{2} = Z_{c}^{2}/\delta Z_{c}^{2}$.
In general this signal-to-noise ratio depends on $\theta_{max}^{2}$
and the experimental parameters $U\cdot\tau$, $\Gamma\cdot\tau$, $\delta k$, and $N$.

We use Eq. (\ref{eq:delta theta^2}) to theoretically estimate $Z_{c}^{2}/\delta Z_{c}^{2}$
and the amplitude sensing limits. We assume the only sources of noise
are projection noise in the measurement of the spin state and fluctuations
in $P_{\uparrow}$ due to the random variation in the relative phase
of the 1D traveling-wave potential and the driven COM motion. Experimentally this is obtained by collecting ~10 photons for each $\ket{\uparrow}$ state, so photon counting shot noise can be neglected \citep{[][{ (see Supplementary Information).}]Bohnet2015}. In this case
$\sigma\left( P_{\uparrow}-P_{\uparrow,bck}\right)=\sqrt{\sigma_{P_{\uparrow}}^2+\sigma_{P_{\uparrow, bck}}^2}$
where the relevant variances are
\begin{equation}
\sigma_{P_{\uparrow, bck}}^2=\frac{1}{N}\left\langle P_{\uparrow}\right\rangle _{bck}\left(1-\left\langle P_{\uparrow}\right\rangle _{bck}\right)=\frac{1}{4N}\left(1-e^{-2\Gamma\tau}\right)\label{eq:Pbck noise}
\end{equation}
and
\begin{equation}
\sigma_{P_{\uparrow}}^2=\sigma_{\delta}^{2}+\frac{1}{N}\left\langle P_{\uparrow}\right\rangle \left(1-\left\langle P_{\uparrow}\right\rangle \right)\:.\label{eq:Pup noise}
\end{equation}
Here $N$ is the number of spins. Equation (\ref{eq:Pbck noise})
and the second term in Eq. (\ref{eq:Pup noise}) are projection noise.
The variance 
\begin{equation}
\sigma_{\delta}^{2}=\left\langle P_{\uparrow}^{2}-\left\langle P_{\uparrow}\right\rangle ^{2}\right\rangle =\frac{e^{-2\Gamma\tau}}{8}\left(1+J_{0}\left(2\theta_{max}\right)-2J_{0}\left(\theta_{max}\right)^{2}\right)\label{eq:phase fluctuation variance}
\end{equation}
is due to the random variation in the relative phase of the 1D traveling-wave and the driven COM motion. For our set-up, $\rm{\it{DWF}}=\exp(-\delta k^2 \left< \hat{z}^{2}_{i} \right> / 2) =0.86$ and
$\delta k=2\pi/\left(900\:\mathrm{nm}\right)$ are fixed, the
decoherence $\Gamma$ is a function of $U$, $\Gamma=\xi\left(U/\hbar\right)$
where $\xi=1.156\times10^{-3}$, and $F_{0} = \rm{\it{DWF}} \cdot U \cdot \delta k$. For a given $Z_{c}$ we use Eqs.
(\ref{eq:theta_max formula}) and (\ref{eq:delta theta^2})-(\ref{eq:phase fluctuation variance})
to find the optimum $Z_{c}^{2}/\delta Z_{c}^{2}$ as
a function of $\left(U\tau\right)/\hbar$. This optimum value is the red dashed
theoretical curve plotted in Fig. 4 of
the main text.

The signal-to-noise $Z_{c}^{2}/\delta Z_{c}^{2}$ is optimized
for relatively small values of $\theta_{max}^{2}$ where $G\left(\theta_{max}^{2}\right)\approx\theta_{max}^{2}/8$
is a good approximation. This leads to some simplifications for Eqs.
(\ref{eq:P_up difference}) and (\ref{eq:delta theta^2}),
\begin{equation}
\theta_{max}^{2}\approx8e^{\Gamma\tau}\left(\left\langle P_{\uparrow}\right\rangle -\left\langle P_{\uparrow}\right\rangle _{bck}\right)\label{eq:theta_max^sq}
\end{equation}
and
\begin{equation}
\delta\theta_{max}^{2}\approx8e^{\Gamma\tau}\sigma\left(P_{\uparrow} - P_{\uparrow,bck}\right)\:,\label{eq:delta theta_max^2}
\end{equation}
and to the following estimate for the signal-to-noise ratio of a single
experimental trial,
\begin{equation}
\frac{\theta_{max}^{2}}{\delta\theta_{max}^{2}} = \frac{Z_{c}^{2}}{\delta Z_{c}^{2}} \approx\frac{\left\langle P_{\uparrow}\right\rangle -\left\langle P_{\uparrow}\right\rangle _{bck}}{\sigma\left(P_{\uparrow} - P_{\uparrow,bck}\right)}\:.\label{eq:S/N limits}
\end{equation}
Figure (4) of the main text uses Eq. (\ref{eq:S/N limits}),
along with repeated measurements of $P_{\uparrow} - P_{\uparrow,bck}$,
to experimentally determine the signal-to-noise ratio as a function
of the imposed amplitude $Z_{c}$ of the COM motion.

Finally we use Eqs. (\ref{eq:theta_max formula}) and (\ref{eq:delta theta^2})-(\ref{eq:phase fluctuation variance})
to calculate the sensing limits for very small $Z_{c}$. For small
$Z_{c}$ the variance $\sigma_{\delta}^{2}$ can be neglected compared
to projection noise and $\sigma_{P_{\uparrow}}^2\approx\sigma_{P_{\uparrow, bck}}^2$.
In this case we obtain the following expression for the signal-to-noise
ratio,
\begin{equation}
\frac{Z_{c}^{2}}{\delta Z_{c}^{2}}=\frac{\sqrt{N}}{4\sqrt{2}}\frac{\rm{\it{DWF}}^{2}\cdot\left(\delta k\,Z_{c}\right)^{2}\left(U\tau/\hbar\right)^{2}}{\sqrt{e^{2\xi U\tau/\hbar}-1}}\:.\label{eq:limiting sensitivity}
\end{equation}
Equation (\ref{eq:limiting sensitivity}) is maximized for $\xi U\tau\approx1.9603$, resulting in 

\begin{equation}
\left.\frac{Z_{c}^{2}}{\delta Z_{c}^{2}}\right|_{\mathrm{limiting}}\approx 0.097 \frac{\sqrt{N}(\rm{\it{DWF}})^2 (\delta k)^2}{\xi^2} Z_c^2\:,
\end{equation}
which is Eq. (5) of the main text.
With $\rm{\it{DWF}}=0.86$, $\delta k=2\pi/\left(900\:\mathrm{nm}\right)$,
$\xi=1.156\times10^{-3}$, and $N=85$,
\begin{equation}
\left.\frac{Z_{c}^{2}}{\delta Z_{c}^{2}}\right|_{optimum}=\left[\frac{Z_{c}}{0.2\:\mathrm{nm}}\right]^{2}\:.\label{eq:26.1 Z_c^2}
\end{equation}
For our set-up and available ODF power, $\xi U\tau/\hbar\approx1.9603$
is realized for $\tau\approx20\:\mathrm{ms}$. A measurement of the
signal and a measurement of the background requires $\sim60\:\mathrm{ms}$, allowing
for $16$ independent measurements of $ P_{\uparrow} - P_{\uparrow,bck}$
in 1 s. The limiting sensitivity is approximately $\left(100\:\mathrm{pm}\right)^{2}$
in a 1 s measurement time, or $\left(100\:\mathrm{pm}\right)^{2}/\sqrt{\mathrm{Hz}}$.
We note that the limiting sensitivity is determined by the ratio $\xi=\Gamma/\left(U/\hbar\right)$.
In particular, the optimum value for Eq. (\ref{eq:limiting sensitivity})
scales as $1/\xi^{2}$.

\section{4. Phase-coherent sensing limits}

With appropriate care the phase of the 1D traveling-wave potential can be stable for long
periods of time with respect to the ion trapping electrodes \citep{Hume2011}, enabling repeated phase-coherent sensing of the same quadrature of the COM motion $Z_{c}\cos(\omega t)$.
In this case the same spin precession $\theta_{max}=\rm{\it{DWF}}\cdot\left(U/\hbar\right)\cdot\delta k\,Z_{c}\cdot\tau$
occurs for each experimental trial, which can be detected to first
order in $\theta_{max}$ (or $Z_{c}$) in a Ramsey sequence with a
$\pi/2$ phase shift between the two $\pi/2$-pulses. Assuming $\sin\left(\theta_{max}\right)\approx\theta_{max}$,
appropriate for small amplitudes $Z_{c}$, the equivalent phase-coherent
sensing expressions for Eqs. (\ref{eq:theta_max^sq}) and (\ref{eq:delta theta_max^2})
are
\begin{equation}
\theta_{max}=2e^{\Gamma\tau}\left(\left\langle P_{\uparrow}\right\rangle -\left\langle P_{\uparrow}\right\rangle _{bck}\right)\label{eq:theta_max coherent}
\end{equation}
and
\begin{equation}
\delta\theta_{max}=2e^{\Gamma\tau}\sigma\left( P_{\uparrow} - P_{\uparrow,bck}\right)\:.\label{eq:delta theta_max coherent}
\end{equation}
For a Ramsey experiment with a $\pi/2$ phase shift, $\left\langle P_{\uparrow}\right\rangle _{bck}=1/2$.
If projection noise is the only source of noise, then for small $Z_{c}$,
$\sigma_{P_{\uparrow}}^2\approx\sigma_{P_{\uparrow, bck}}^2=\frac{1}{N}\cdot\frac{1}{2}\cdot\frac{1}{2}$
and $\sigma\left( P_{\uparrow} - P_{\uparrow,bck} \right)\approx\frac{1}{\sqrt{2N}}$.
The limiting signal-to-noise ratio $\theta_{max}/\delta\theta_{max}$
of a $(P_{\uparrow} - P_{\uparrow,bck})$
measurement is
\begin{equation}
\frac{\theta_{max}}{\delta\theta_{max}}=\frac{Z_{c}}{\delta Z_{c}}=\rm{\it{DWF}}\cdot\left(\delta k\,Z_{c}\right)\cdot\sqrt{\frac{N}{2}}\cdot\frac{\left(U\tau\right)}{\hbar}e^{-\xi U\tau/\hbar}\:.\label{eq:S/N coherent}
\end{equation}
Equation (\ref{eq:S/N coherent}) is maximized for $\xi U\tau/\hbar=1$.
With $\rm{\it{DWF}}=0.86$, $\delta k=2\pi/\left(900\:\mathrm{nm}\right)$,
$\xi=1.156\times10^{-3}$, and $N=100$,
\begin{equation}
\left.\frac{Z_{c}}{\delta Z_{c}}\right|_{optimum}=\frac{Z_{c}}{0.074\:\mathrm{nm}}\:.\label{eq:coherent optimum}
\end{equation}
With $16$ independent measurements of $\left\langle P_{\uparrow}\right\rangle -\left\langle P_{\uparrow}\right\rangle _{bck}$
in 1 s, this corresponds to a limiting sensitivity of $\sim\left(20\:\mathrm{pm}\right)/\sqrt{\mathrm{Hz}}$.
The optimum value for the signal-to-ratio of Eq. (\ref{eq:S/N coherent})
scales as $1/\xi$. By employing spin-squeezed states that
have been demonstrated in this system \citep{Bohnet2015},
$Z_{c}/\delta Z_{c}$ can be improved by another factor
of 2.

Employing this technique to sense motion on resonance with the COM mode
can lead to the detection of very weak forces and electric fields.
The detection of a 20 pm amplitude resulting from a 100 ms coherent
drive on the 1.57 MHz COM mode is
sensitive to a force/ion of $5 \times 10^{-5}\:\mathrm{yN}$, corresponding
to an electric field of $0.35\:\mathrm{nV/m}$.

\bibliographystyle{apsrev4-1}
\bibliography{amplitude_sensing}